\documentclass[fleqn,twoside]{article}
\usepackage{espcrc2}

\usepackage{graphicx}
\usepackage[figuresright]{rotating}

\newcommand{\ie}{{\it i.e.}}

\newcommand{\qu}{{\rm q}}

\newcommand{\qbm}{{\rm\bar q}}

\newcommand{\rvec}{\vec r}
\newcommand{\Rvec}{\vec R}

\newcommand{\ieps}{i\varepsilon}

\newcommand{\pl}{{||}}

\newcommand{\eq}[1]{(\ref{#1})}

\newcommand{\beq}{\begin{equation}}
\newcommand{\eeq}{\end{equation}}
\newcommand{\beqa}{\begin{eqnarray}}
\newcommand{\eeqa}{\end{eqnarray}}

\newcommand{\PL}[3]{Phys.~Lett.~{\bf {#1}},~{#2}~({#3})}
\newcommand{\NP}[3]{Nucl.~Phys.~{\bf {#1}},~{#2}~({#3})}
\newcommand{\PRD}[3]{Phys.~Rev.~{\bf D{#1}},~{#2}~({#3})}

\newcommand{\PRe}[3]{Phys.~Rep.~{\bf {#1}},~{#2}~({#3})}

\newcommand{\AmS}{{\protect\the\textfont2
  A\kern-.1667em\lower.5ex\hbox{M}\kern-.125emS}}

\title{\large STRUCTURE FUNCTIONS ARE NOT PARTON PROBABILITIES}

\author{S. Peign\'e\address{LAPTH, BP 110, F-74941 Annecy-le-Vieux
    Cedex, France},
F. Sannino\address{NORDITA, Blegdamsvej 17, DK--2100 Copenhagen,
Denmark}}

\begin{document}

\begin{abstract}
We explain why contrary to common belief, the deep
inelastic scattering structure functions are not related to
parton probabilities in the target. \vskip 5pt
(Talk presented during the `International Light-Cone
Workshop', Trento, ECT$^*$, September 3-11, 2001.)
\end{abstract}
\maketitle

Our statement originates from
noting the incompatibility between the
Glauber-Gribov picture of nuclear shadowing and the deep
inelastic scattering (DIS) cross section $\sigma_{DIS}$ being
determined by parton pro\-ba\-bi\-li\-ties.
In section 1 we present physical arguments for this incompatibility.
In section 2 we prove our main statement by an explicit
calculation in a scalar QED model.

Our talk is based on the work done in \cite{bhmps}, to which we
refer for technical details.

\section{PARTON DISTRIBUTIONS VERSUS GLAUBER-GRIBOV NUCLEAR SHADOWING}

According to the QCD factorization theorem \cite{css},
$\sigma_{DIS}$ can be expressed, at leading-twist and in the
Bjorken limit\footnote{Choosing $q=(\nu, \vec{0}_{\perp}, q^z)$
and $p=(M, \vec{0}_{\perp}, 0)$ for the momenta of the virtual
photon and of the target, the Bjorken limit corresponds to
$Q^2=-q^2 \to \infty$ and $\nu \to \infty$ with
$x_B=Q^2/(2M\nu)=Q^2/M(q^-+q^+)$ being fixed. We use the
light-cone variables $q^{\pm}=q^0\pm q^z$.}, as a convolution
between hard partonic subprocess cross sections and parton
distributions.  The quark distribution in the nucleon $N$ of
momentum $p$ reads
\beqa \label{fq}
f_{\qu/N}(x_B,Q^2)= \int \frac{dy^-}{8\pi} e^{-ix_B p^+ y^-}
 \times \ \ \ \ \ \  \nonumber \\ \langle N| \qbm(y^-) \gamma^+ {\rm
  P}\exp\left[ig\int_0^{y^-}dw^{-} A^+(w^-)  \right]
\qu(0)|N\rangle  \\ \nonumber
\eeqa
Note that the (frame-independent) expression for $f_{\qu/N}$
is written in \eq{fq} with 
a particular choice of Lorentz frame,
namely $q^z <0$. In the Bjorken limit we then have $q^- \simeq 2
\nu \gg q^+$ and $q^+ = -Q^2/q^- \simeq - Mx_B <0$. In such a $q^+
<0$ frame, the virtual photon fluctuation $\gamma^* \to \qu \bar
\qu$ is forbidden in light-cone time-ordered (LCTO) perturbation
theory. Hence only LCTO diagrams where $\gamma^*$ is {\it
absorbed} can contribute. In this framework initial (ISI) and
final (FSI) state interactions are then defined as occurring
before or after the $\gamma^*$ absorption time.

Gauge invariance of the matrix element in \eq{fq} is ensured by
the presence of the path-ordered exponential. The latter is
responsible, in Feynman gauge and more generally in covariant
gauges, for FSI of the struck quark $p_1$ (see Fig. 1) which
affect $\sigma_{DIS}$. The path-ordered exponential reduces to the
identity in the light-cone $A^+=0$ gauge. Thus if \eq{fq} is
correct also in this particular gauge, as usually believed,
$f_{\qu/N}$ is then given by the square of the target nucleon wave
function (evaluated in $A^+=0$ gauge) \cite{spd}. This
would allow a probabilistic interpretation of the structure functions.
In other words, only ISI would contribute to $\sigma_{DIS}$ in $A^+=0$ gauge.

We will demonstrate in section 2, in the framework of a simple
model, that FSI do modify $\sigma_{DIS}$ in {\it all} gauges,
including the $A^+=0$ gauge. Hence $f_{\qu/N}$ cannot be simply
interpreted as a parton probability and its expression \eq{fq},
although valid in covariant gauges, does not actually hold in
$A^+=0$ gauge. The fact that this gauge is plagued with
singularities has already been pointed out in \cite{css}.
Our results suggest that a more general expression for $f_{q/N}$
taking into account the $A_\perp$ components
of the target gauge field is needed.
These components might formally be included by writing
the path-ordered exponential in the covariant form ${\rm
P}\exp\left[ig\int_0^{y} dw_\mu A^\mu(w) \right]$.

The Glauber-Gribov picture of nuclear shadowing \cite{gribov,pw}
illustrates why structure functions and parton probabilities
cannot be simply related. Roughly speaking the shadowing-type
contribution is encoded in the structure functions and arises as a
quantum interference effect between (in general {\it complex})
rescattering amplitudes while parton probabilities are related
only to the {\it real} wave function of a stable target.

More specifically in a Lorentz frame where $q^+>0$, the $\gamma^*
\to \qu \bar \qu$ fluctuation is possible in a LCTO formulation
and occurs long before the target for $x_B\ll 1$.
An interference can arise between an amplitude where
the $\qu \bar \qu$ pair scatters inelastically on a target nucleon
$N_2$ and another where it undergoes first an elastic scattering
(via Pomeron exchange) on a nucleon $N_1$ (see Fig. 1). At small
$x_B$ the interference is destructive because the exchanged
Pomeron amplitude becomes purely imaginary while the intermediate
state after the elastic scattering is quasi on-shell. But note
that the presence of on-shell intermediate states (OSIS) is
independent of the Lorentz frame {\it and} of the gauge. Hence the
Glauber-Gribov mechanism viewed in the frame of interest $q^+ <0$
(where the quark distribution \eq{fq} is expressed) and in $A^+=0$
gauge must also involve OSIS. These OSIS cannot arise from ISI
(they would constitute decay channels of the target) but from FSI.
Since shadowing affects $\sigma_{DIS}$, this suggests that FSI
modify $\sigma_{DIS}$ (\ie \ do not cancel when summing over cuts
in the forward DIS amplitude) in all gauges including $A^+=0$.
This contradicts the probabilistic interpretation of parton
distributions, according to which in the $A^+=0$ gauge only ISI,
which build the wave function, can influence the DIS cross
section.
\begin{figure}[htb]
\includegraphics[width=18pc]{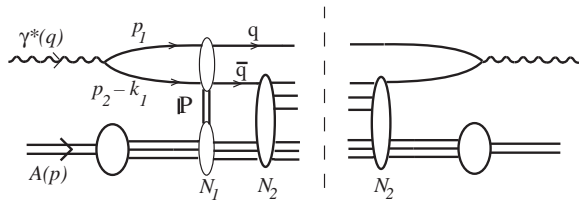}
\caption{Glauber-Gribov shadowing involves interference between
rescattering amplitudes.}
\label{fig2dis}
\end{figure}

\section{A PERTURBATIVE SCALAR QED MODEL FOR RESCATTERING}

We now present a toy model which captures the main features of the
shadowing phenomenon. Since the color and spin degrees of freedom
are not essential for our conclusions, we consider scalar QED
and compute the specific contribution to the forward DIS
amplitude depicted in Fig. 2. As target $T$, we take a scalar
`heavy quark' of mass $M$. The incoming virtual photon couples
with charge $e$ to a scalar `light quark' of mass $m$. The heavy
and light quarks interact via `gluon' (photon) exchanges with coupling $g$.
As we will see, the particular contribution of Fig. 2 involving four
`gluon' exchanges exhibits the features of
shadowing mentioned in section 1. The square of the Pomeron
exchange amplitude is obtained when cutting the diagram
between the exchanges $k_2$ and $k_3$ while cutting between
$k_3$ and $k_4$ gives the interference of Fig. 1 (two of the
three gluons of the left amplitude then model a
Pomeron). The contribution of the diagram shown in Fig. 2
to $\sigma_{DIS}$ is negative (see \eq{sumofcuts})
and thus reduces the
Born cross section (corresponding to two exchanged gluons instead
of four).
\begin{figure}[htb]
\includegraphics[width=18pc]{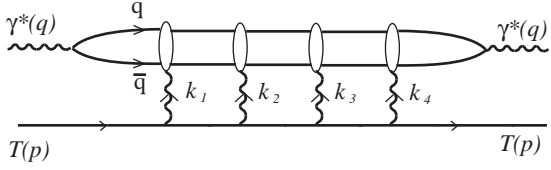}
\caption{Forward $\gamma^* T \to \gamma^* T$ amplitude. All
attachments of  the exchanged gluons to the upper scalar loop are included,
as well as topologically distinct permutations of the lower vertices on the
target line.}
\label{fig3dis}
\end{figure}

Let us note that since we consider an elementary target, our model
is not meant to really describe nuclear shadowing, but is however
ade\-qua\-te to study the effect of FSI on the cross section. Our
perturbative treatment allows to study precisely the structure of
what would be called the `soft' part of the dynamics in QCD
factorization theorems \cite{css}. We take the hard part at zeroth
order in g and focus on the aligned jet kinematics (where the
`hard' quark $p_1$ takes all the incoming photon energy whereas
the antiquark $p_2$ remains `soft') in the Bjorken limit. We then
consider the small $x_B$ limit both for simplicity and to recover
features of shadowing. Our final statements however do not
depend on the small $x_B$ limit. We consider the kinematics: \beq
2\nu \sim p_1^- \gg p_2^- \gg k_{i\perp},\ p_{i\perp},\ k_i^-, \ m
\gg k_i^+ \label{Mscales} \eeq In order to evaluate the sum of the
three re\-le\-vant cuts of the forward $\gamma^*(q) T(p) \to
\gamma^*(q) T(p)$ amplitude, we need to calculate three basic
ingredients, the amplitudes $\gamma^*(q) T(p) \to \qu(p_1) \bar
\qu(p_2)  T(p')$ corresponding to one-, two- and three-gluon
exchange which we call $A$ , $B$ and $C$ respectively. Typical
diagrams contributing to $B$ in the small $x_B$ limit and in
Feynman gauge are shown in Fig. 3 (see \cite{bhmps} for more
details).

These amplitudes can be expressed in the transverse coordinate
space ($\rvec_\perp$ denotes the transverse separation between the
quark $p_1$ and antiquark $p_2$, and $\Rvec_\perp$ the distance
between the target quark and the light $\qu \bar \qu$ pair)
\cite{bhmps}: \beqa A &=& 2eg^2 M Q p_2^-\,
V(m_\pl r_\perp) W(\rvec_\perp, \Rvec_\perp) \nonumber \\
B &=&  \frac{-ig^2}{2!} W A \nonumber \\
C &=& \frac{(-ig^2)^2}{3!} W^2 A
\label{ABC}
\eeqa
where
\beqa
m_\pl^2 &=& p_2^-Mx_B + m^2 \nonumber \\
V(m\, r_\perp) &=& \frac{1}{2\pi}K_0(m\,r_\perp) \nonumber \\
W(\rvec_\perp, \Rvec_\perp)  &=& \frac{1}{2\pi}
\log\left(\frac{|\Rvec_\perp+\rvec_\perp|}{R_\perp} \right)
\eeqa
The function $V$ denotes the incoming longitudinal photon wave
function. In the case of scalar quarks, the longitudinal photon
polarization indeed dominates in the Bjorken limit.
The function $W$ stands for the
amplitude associated to dipole scattering. Thus $A \propto W$, $B
\propto W^2$ and $C \propto W^3$. Our results \eq{ABC} are gauge
independent\footnote{We obtained \eq{ABC} in both Feynman and
$A^+=0$ gauges. In $A^+=0$ gauge, we showed the result to be
independent of the prescription used to regularize the spurious
$k_i^+=0$ poles \cite{bhmps}.}. Note that $B$ is imaginary in the
small $x_B$ limit and $B \propto p_2^-$, hence $B$ incorporates
the main features of Pomeron exchange.
\begin{figure}[htb]
\includegraphics[width=18pc]{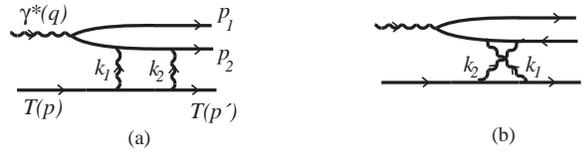}
\caption{Diagrams which give leading order contributions to the
one-loop amplitude $B$ in Feynman gauge.}
\label{fig5dis}
\end{figure}

The contribution of Fig. 2 to $\sigma_{DIS}$ is
\beqa
\Delta \left( Q^4\frac{d\sigma}{dQ^2\, dx_B} \right) =
\frac{\alpha}{16\pi^2}\frac{1-y}{y^2}
\frac{1}{2M\nu} \int \frac{dp_2^-}{p_2^-} &&\nonumber \\
\times \int d^2\rvec_\perp\, d^2\Rvec_\perp\,
\left[ |B|^2 + 2 A C \right] &&
\label{deltasigmaDIS}
\eeqa
This is easily seen to be non-zero and negative:
\beq
|B|^2 + 2 A C = - \frac{1}{3} \frac{g^4}{4} A^2 W^2 <0
\label{sumofcuts}
\eeq
so that \eq{deltasigmaDIS} gives a negative correction
to the Born DIS cross section given by $A^2$, as expected for
shadowing. The interference term $\sim 2 A C$ overcompensates the
$|B|^2$ term.

The rescattering correction to $\sigma_{DIS}$ is leading-twist
since $|B|^2 + 2 A C \propto Q^2$, which makes the r.h.s. of
\eq{deltasigmaDIS} independent of $Q$ in the Bjorken limit.

In Feynman gauge it can be shown that the partial contribution to
\eq{deltasigmaDIS} from
rescatterings of the `soft' antiquark $p_2$ vanishes\footnote{This is also
true in a general covariant gauge, but not in $A^+=0$ gauge. Indeed
the subset of diagrams involving only rescatterings of $p_2$ is
gauge dependent.}. Thus only rescatterings of the hard quark $p_1$
contribute to \eq{sumofcuts} in Feynman gauge. This is nothing else
than the confirmation of the presence of FSI in covariant gauges.

Now comes our central argument. At small $x_B$ the two-gluon
exchange amplitude $B$ $\gamma^*(q) T(p) \to \qu(p_1) \bar
\qu(p_2)  T(p')$ arises from the intermediate state occurring
between the first and second/last gluon exchange being on-shell.
(This intermediate state becomes exactly on-shell in the $x_B \to
0$ limit.) The typical LC
distance $y^-$ between the exchanges behaves as $\sim 1/x_B \to
\infty$. This can be seen by calculating $B$ in LCTO perturbation
theory, where every intermediate state is associated with a
denominator factor
\beq D_{int}=\sum_{inc}k^- - \sum_{int}k^-
+\ieps \label{dint}
\eeq
which measures the LC energy difference
between the incoming and intermediate states. An intermediate
state is on-shell when only $\sum_{inc}k^- - \sum_{int}k^- = 0$
contributes to the amplitude.
In the matrix element (1), all rescatterings occur at equal time
$y^+$, whereas the LC distance $y^-\sim 1/x_B$.
Both our amplitudes $B$ and $C$
arise from such OSIS in the $x_B \to 0$ limit. (An infinite
distance is allowed between any of the gluon exchanges.)
Obviously, the expression $|B|^2 + 2 A C$, built from $B$ and $C$,
also arises purely from OSIS. The pre\-sen\-ce of OSIS being gauge
independent, we conclude that $B$ and $C$, and thus $|B|^2 + 2 A
C$, originate from OSIS also in $A^+=0$ gauge. As stated previously,
these OSIS cannot arise from ISI for a stable target. They can be
created only via FSI. We conclude:

{\it In any gauge FSI are present and do modify the DIS cross
  section. Thus structure functions cannot be simply related to parton probabilities.}

\vspace{.5cm}

{\bf Acknowledgements.} It is a pleasure for us to thank S.
Brodsky, P. Hoyer and N. Marchal for a very stimulating collaboration.


\vspace{1cm}




\begin{thebibliography}{9}

\bibitem{bhmps}
S. Brodsky, P. Hoyer, N. Marchal, S. Peign\'e and
F. Sannino, hep-ph/0104291, submitted to Phys.Rev.D.

\bibitem{css}
J. C. Collins and D. E. Soper, \NP{B194}{445}{1982};\\
J. C. Collins, D. E. Soper and G. Sterman, \NP{B261}{104}{1985},
\NP{B308}{833}{1988}, \PL{B438}{184}{1998} and review in {\it Perturbative
Quantum Chromodynamics}, (A.H. Mueller, ed.,World Scientific Publ., 1989, pp.
1-91); \\
G. T. Bodwin, \PRD{31}{2616}{1985}, Erratum \PRD{34}{3932}{1986}.

\bibitem{spd}
S. J. Brodsky, M. Diehl and D. S. Hwang, \NP{B596}{99}{2001} [hep-ph/0009254]; \\
M. Diehl, T. Feldmann, R. Jakob and P. Kroll, \NP{B596}{33}{2001}
[hep-ph/0009255].

\bibitem{gribov}
V. N. Gribov, Sov. Phys. JETP {\bf 29}, 483 (1969) and {\bf 30}, 709 (1970).

\bibitem{pw}
G. Piller and W. Weise, \PRe{330}{1}{2000} [hep-ph/9908230]; \\
G. Piller, M. V\"anttinen, L. Mankiewicz and W. Weise, hep-ph/0010037.

\end{thebibliography}
\end{document}